\documentclass{optica-article}

\journal{opticajournal} % for journals or Optica Open

\articletype{Research Article}

\usepackage{lineno}
\usepackage{physics}
\usepackage{cleveref}
\usepackage{hyperref}
\usepackage{wallpaper}
\usepackage{tabularx} 
\usepackage{hyperref}

\begin{document}

\title{Ultra-precise phase estimation without mode entanglement}

\author{Mikhail S.Podoshvedov,\authormark{1,2} Sergey A. Podoshvedov,\authormark{1,2,*}}

\address{\authormark{1}Laboratory of quantum information processing and quantum computing, South Ural State University (SUSU), Lenin Av. 76, Chelyabinsk, Russia\\
\authormark{2}Laboratory of quantum engineering of light, South Ural State University (SUSU), Lenin Av. 76, Chelyabinsk, Russia}

\email{\authormark{*}sapodo68@gmail.com} %% email address is required; see note below about the corresponding author designation

% use {asbstract*} to suppress the copyright line. Copyright information will be added in production

\begin{abstract*} 
We explore optical quantum engineering of phase-parameterized continuous-variable ($CV$) probe states to exploit nonclassical light to solve the problem of precise phase estimation. The optical interferometer consists of a single beam splitter ($BS$) with tunable transmittance and reflectance, and two single-mode squeezed vacuum states ($SMSVs$). The reference $SMSV$ state is mixed with a weakly squeezed state carrying an unknown phase at the beam splitter to form an output hybrid entangled state. Then, in the measurement mode, the number of photons is measured to generate the target $CV$ state parameterized by the unknown phase. Using the $CV$ states, we propose a sub-Heisenberg metrology protocol in which the quantum Cramer-Rao ($QCR$) boundary is saturated by intensity measurement. The advantage of quantum engineering of $CV$ probe states for ultra-precise phase estimation of unknown phase is due solely to the nonclassical photonic properties of the measurement induced $CV$ states of definite parity and is independent of the mode entanglement.     

\end{abstract*}

%%%%%%%%%%%%%%%%%%%%%%%%%%  body  %%%%%%%%%%%%%%%%%%%%%%%%%%
\section{Introduction}
The ability to obtain an extremely accurate estimate of an unknown parameter is a fundamental requirement of most scientific research. Examples of the precision measurements involve measurement of weak gravitational signals via optical interferometer \cite{chap1:1} \cite{2}, the precise measurement of transition frequencies of atoms and molecules \cite{3}, and the fabrication of nanodevices using optical lithography \cite{4}. Therefore, one of the main goals of researchers is to develop a method for extracting as much information as possible about unknown parameters from measurement data. Quantum metrology utilizes quantum states to achieve greater sensitivity in phase shift measurements than is possible with classical methods alone, which at best can provide sensitivity at the level of shot noise or the standard quantum limit ($SQL$) \cite{5}\cite{6}\cite{7}\cite{8}. If the interferometer operates with a classical light source, then the best sensitivity scales as the inverse of the square root of the average number of photons in the state being used. Light without a classical analog can overcome the $SQL$ sensitivity limit at the interferometer output. This is achieved by exploiting properties inherent to the quantum state, such as squeezing \cite{10}\cite{11}\cite{12}\cite{13}\cite{14}\cite{15} and/or entanglement \cite{16}\cite{17}. 
\par Among the popular interferometric platforms, the Mach-Zehnder ($MZ$) interferometer and SU(1,1) interferometer are worth mentioning. A simple $MZ$ interferometer design \cite{18} allows for obtaining the corresponding results in three stages. The first beam splitter prepares the probe state, the sample determines the interaction by setting the unknown parameter and second $BS$ photon detectors represent the measurement stage. The phase uncertainty of coherent states directed to the $MZ$ interferometer input and estimated from the intensity measurement results reaches the $SQL$. This is not surprising, since the Poisson statistics of coherent states arise from the independence of events. To further reduce the phase uncertainty at the quantum level of the Heisenberg limit ($HL$), appropriate engineering of high-intensity non-classical sources is required \cite{19}\cite{20}. In SU interferometer proposed in \cite{21} passive optical elements are replaced by optical parametric amplifiers (OPA). Despite the significant potential of the SU interferometers for precision measurements \cite{22}\cite{23}\cite{24}\cite{25}, their practical effectiveness is limited by photon losses \cite{26}\cite{27}, which is a serious obstacle to their development.
\par In general, any protocol of quantum estimate of unknown parameter can be divided into three distinct sections: preparing the probe state, parameterizing it through interaction with the corresponding system, and measurement of the output state to estimate the unknown parameter \cite{7}\cite{8}. All three stages are interconnected in achieving the final goal, and deficiencies in any one step can impair or even compromise the quantum mechanical approach. In most cases, a special observable is required to overcome $SQL$ even in the presence of a highly nonclassical state. For example, the quantum Fisher information ($QFI$) of the $SMSV$ state, which determines $QCR$ boundary \cite{28} of the phase-parametrized state being ultimate precision of the state regardless of the observable used, reaches the value $F_{SMSV}=4\triangle_{SMSV}^2=8(\langle{n_{SMSV}}\rangle^2+\langle{n_{SMSV}}\rangle)$, where $\triangle_{SMSV}^2$ and $\langle{n_{SMSV}}\rangle$ are the variance and average number of photons in the $SMSV$ state. Therefore, the ultimate phase uncertainty of the $SMSV$ state can potentially be estimated with sub-Heisenberg precision, but its intensity is insensitive to phase changes, so its measurement is carried out with a large error. In general, not all observables are able to utilize the full potential of the nonclassical states to reach $QCR$ boundary. Measuring the intensity difference at the output of the $MZ$ interferometer often introduces the phase uncertainty significantly exceeding the $QCR$ limit \cite{13}\cite{15}. Measuring the parity of the number of photons at the output of the interferometer may even prove more useful in practice \cite{12}\cite{29}. For required parameter, the choice of the probe state and corresponding observable is a strategy of the parameter estimate.      
\par All this motivates to the development of high-precision quantum metrological strategies that encompass all stages, starting with the quantum engineering of light aimed at creating probing states to the selection of a suitable observable whose measurement could fully reveal the nonclassical properties of the probe states. Here we develop feasible approach from quantum engineering of new nonclassical light to its measurement in order to achieve the minimum possible error in estimating the unknown phase. Mixing two $SMSV$ states, one reference and the other weakly squeezed with an unknown phase, on a beam splitter with arbitrary transmission and reflection, followed by recording the number of photons in the measurement mode, generates the probe phase-parameterized state $CV$ state of definite parity. The inclusion of beam splitters with arbitrary parameters allows for an expansion of the applicability of the basic optical elements, which is reflected in the study of nonclassical properties of the measurement induced macroscopic $CV$ states of a certain parity \cite{30}, as well as in an increase of the squeezing in photon subtracted $CV$ states \cite{13}. Furthermore, the expanded capabilities of the beam splitter lead to an increase in the phase sensitivity of the $MZ$ interferometer when detecting parity \cite{31}, enables the generation of large amplitude even/odd superposition of coherent states $SCSs$ with high fidelity \cite{32}, and provide a nearly deterministic transfer of entanglement from a nonlocal photon to the initially separated $SMSV$ states \cite{33}. The output probe state, parameterized by the unknown phase, is a superposition of two $CV$ states of a certain parity with a nonzero overlap, allowing its $CV$ components to interfere to interfere in the intensity measurement. The simplest detection scheme using a photodiode is the intensity measurement, being saturating for the $CV$ probe state under study, that is, the phase uncertainty estimated by the error propagation formula becomes almost equal to the $QCR$ boundary. Moreover, measuring the intensity of the $CV$ stares allows us to achieve sub-Heisenberg level of precision. Genuine photon-number resolution can be achieved using a superconducting transition edge sensor $(TES)$ \cite{34}\cite{35}, which is a bolometer maintained near the transition temperature. Given the small number of subtracted photons (from 1 to 4), the strategy with $CV$ states of definite parity is feasible in practice and, moreover, quite stable with respect to the quantum efficiency of the $TES$ detector.

\section{QCR boundary and phase error in intensity measurement of phase-parameterized probe CV states}

Photons can serve as ideal probes for the ultra-precise estimate of an unknown parameter. Indeed, phase information can be encoded in photonic states using their various degrees of freedom \cite{8}\cite{28}\cite{29}. The path encoding can be considered as most widely used photonic degree in quantum metrology. The latter refers to the set of spatial modes occupied by a photon. To fully control such photons, it is sufficient to have beam splitters and phase shifters, the advantage of which can be completely demonstrated in interferometric schemes \cite{18}. In general, the interferometer is a device capable of transforming an input probe state $\rho$ in a such way that output state $\rho_{\varphi}$ can be parameterized by real unknown parameter, say $\varphi$. Therefore, interferometric setups, say $MZ$ interferometer, are the basis of most quantum metrology tasks \cite{5}\cite{6}\cite{7}. At the end of a process, it is necessary to measure the probe state with information encoded in it and, based on the measurement results, estimate the parameter $\varphi$. Measuring the intensity difference is a standard for optical interferometry with coherent states. There are, however, disadvantages to using the interferometry. Typical phase dependent methods suffer from a limited dynamic range, since light fields with phases differing by $2\pi$ can correspond to the same measurement output signal. The $MZ$ interferometry is sensitive to changes in the relative phase between two arms, meaning that such a sensor can require potentially impractical or expensive stabilization, as is typical for classical interferometry.

\begin{figure}[htbp]
    \centering\includegraphics[trim={1cm 3cm .2cm 3.5cm},clip]{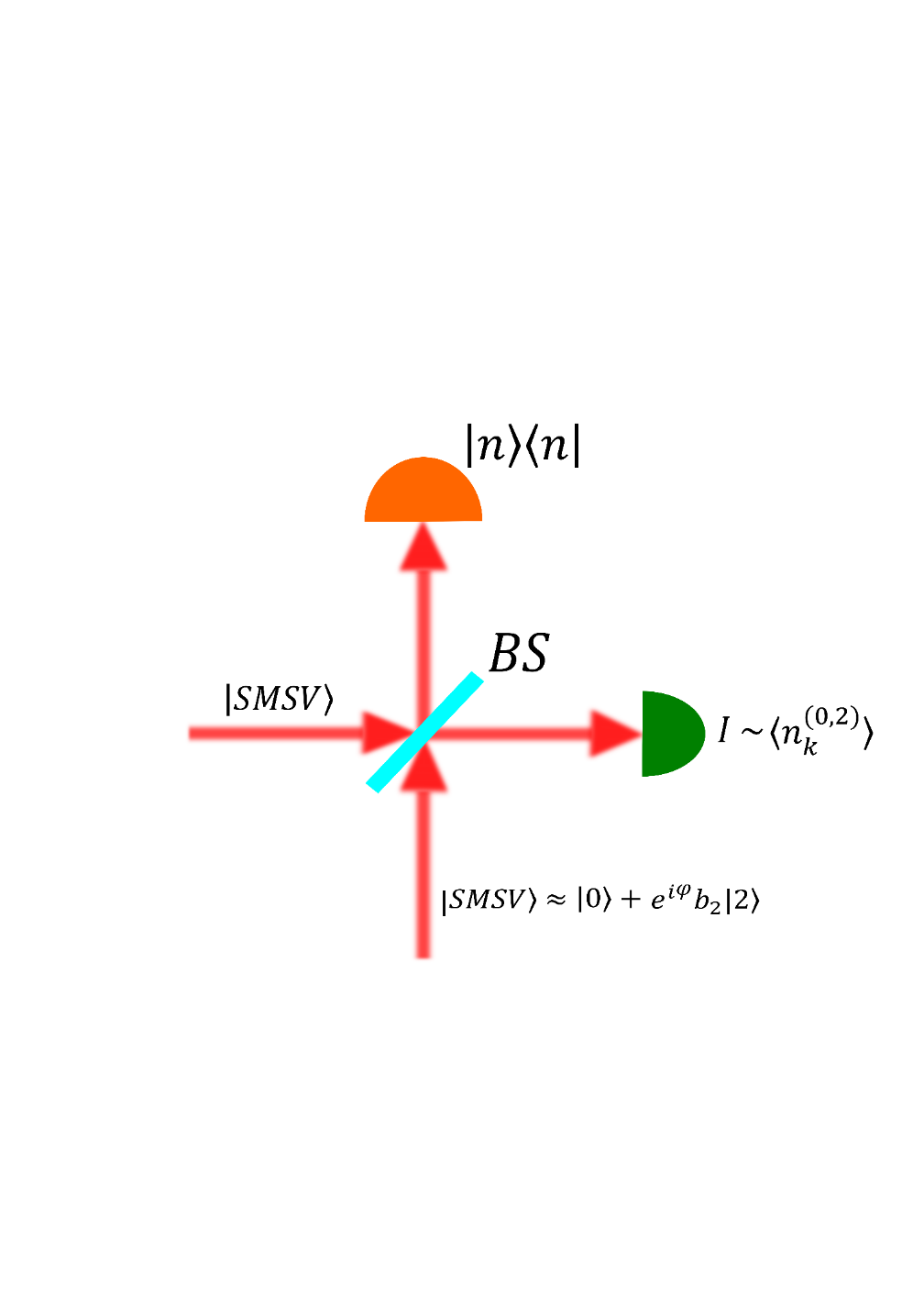}
    \caption{Schematic representation of the protocol for quantum estimation of unknown phase shift $\varphi$ using reference and auxiliary $SMSV$ states. The auxiliary $SMSV$ state contains an unknown parameter, and its mixing with the reference $CV$ state on a beam splitter with arbitrary real transmittance $t>0$ and reflectance $r>0$  is the basis for quantum engineering of the $CV$ state of definite parity. After measuring a certain number of photons in the measurement mode using the $PNR$ detector, initial probe $SMSV$ state is parametrized by unknown parameter $\varphi$. In the remaining mode the average number of photons is measured to estimate the unknown parameter $\varphi$ with sub-Heisenberg precision.}
    \label{Figure.1}
\end{figure}

\par Typical $CV$ states, such as the $SMSV$ state, do not have the advantage of determining the unknown phase parameter from the signal, since the Fock basis states are orthogonal to each other. Direct measurement of the intensity of the probe state parameterized by an unknown phase parameter is not sensitive to phase change. To determine the unknown parameter by direct measurement without $MZ$ interferometry, it is necessary to create a new probe $CV$ state with information encoded in its non-orthogonal components. Here, we start the method of quantum engineering of the probe $CV$ state with consideration of the $SMSV$ state as the reference light state in \hyperref[Figure.1]{Figure.~\ref*{Figure.1}}

\begin{equation}\label{eq:1}
    \ket{SMSV(y)} =
    \frac{1}{\sqrt{\cosh{s}}}\sum_{n=0}^{\infty} \frac{y^{n}}{\sqrt{(2n)!}}\frac{(2n)!}{n!}\ket{2n}  
\end{equation}

\!\!\!\!\!\!the canonical form \cite{30} of which is represented by replacing the real squeezing parameter $0\leq y\leq0.5$ on the squeezing amplitude $y=\tanh{s}⁄2$ with $s>0$. The squeezing can also be expressed in decibels $S=-10 \lg{(\exp{(-2s)})}$ $dB$. Note the mean number of photons in the $SMSV$ state is $\langle n_{SMSV} \rangle= \sinh ^2 s$.
As an auxiliary $CV$ state we also use the $SMSV$ state with a much smaller squeezing amplitude $s_{2} \ll 1$, which allows us to approximate it with a superposition of the vacuum and the two-photon state

\par As an auxiliary $CV$ state we also use the $SMSV$ state with a much smaller squeezing amplitude $s_2\ll1$, which allows us to approximate it with a superposition of the vacuum and the two-photon state

\begin{equation}\label{eq:2}
    \ket{SMSV(y_2)} \approx \ket{\xi} = 
    \frac{1}{\sqrt{N_2}}(\ket{0} + \exp(i\varphi)b_2\ket{2}),  
\end{equation}

\!\!\!\!\!\!where $b_2 = \tanh s_2 / \sqrt 2$ and $N_2=1+b_2^2$ being the corresponding normalization factor. 
The unknown phase shift $\varphi$, which must be estimated, and the amplitude of the two-photon state follows directly from the definition of the $SMSV$ state \cite{30}\cite{32}. This formulation of the problem with two $CV$ states \eqref{eq:1} and \eqref{eq:2} is practical since the $SMSV$ states can be realized deterministically using nonlinear optics. The most successful method uses degenerate optical parametric down conversion of type $I$ both inside the optical resonator and even during a single pump pass through the crystal \cite{36}.
\par Mixing the reference state in Eq. \eqref{eq:1} with the superposition $\ket{\xi}$ on the beam splitter with arbitrary real transmission $t>0$ and reflection $r>0$ amplitudes followed by measuring $k$ photons in the measurement mode, as shown in \hyperref[Figure.1]{Figure.\!~\ref*{Figure.1}}, allows generating new measurement induced $CV$ states of a certain parity

\begin{equation}\label{eq:3}
    \ket{\Psi_k^{(02)}(y_1,B,\varphi)}  = 
    \frac{1}{\sqrt{G_k^{(02)}(y_1,B,\varphi)}}\Bigl(\ket{\Psi_k^{(0)}(y_1)} + \exp(i\varphi)b_2b_k^{(2)}\ket{\Psi_k^{(2)}(y_1,B)}\Bigl),  
\end{equation}

\!\!\!\!\!\!the details of the derivation of which are presented in the supplementary material, where its additional factor $b_{k}^{(2)}$ is given by

\begin{equation}\label{eq:4}
    b_{k_1k_2}(y_1,B) = \frac{1}{\sqrt{2}(1 + B)}
    \begin{cases}
            \frac{B}{y_1}\sqrt{\frac{G_0^{(2)}(y_1,B)}{Z(y_1)}}, 
            \,\,\text{if}\,\,\, k = 0 \\[3ex]
			-\frac{2}{y_1}\sqrt{\frac{G_1^{(2)}(y_1,B)}{Z^{(1)}(y_1)}}, & 
            \!\!\!\!\!\!\!\!\!\!\!\!\!\!\!\!\!\!\!\!\!\text{if } k = 1 \\[3ex]
			\frac{k(k-1)}{y_1B}\sqrt{\frac{G_{k}^{(2)}(y_1,B)}{Z^{(k)}(y_1)}}, &\!\!\!\!\!\!\!\!\!\!\!\!\!\!\!\!\!\!\text{if } k > 1 
	\end{cases}
\end{equation}

\!\!\!\!\!\!The additional multiplier follows directly from the difference in the interaction of the $SMSV$ state with the vacuum and the two-photon state and it changes the initial amplitude $b_2$ of the superposition \eqref{eq:2} by $b_k^{(2)}$ ) times. 
\par The derivation of the $CV$ state in equation \eqref{eq:3} uses a realistic approach to the interaction of the $SMSV$ state with photonic states on an arbitrary $BS$, the details of which are presented in the supplementary materials. Including a beam splitter with variable transmittance and reflectance allows one to consider the problem not only for highly transmitting $BS$, but also for balanced one, as well as for those that redirect a significant part of the energy from the reference wave to the measuring mode, i.e., highly reflective $BSs$ \cite{32}\cite{33}. 
The final superposition \eqref{eq:3} consists of two states of definite parity $\ket{\Psi_{k}^{(0)}(y_1)}$  and $\ket{\Psi_{k}^{(2)}(y_1,B)}$ , in which the superscript either (0) or (2) indicates the number of additional input photons, and the subscript $k$ is responsible for the number of subtracted, i.e. measured in the auxiliary measurement mode, photons. If $k=2m$ is even then the superposition \eqref{eq:3} is also even, that is, it consists of even Fock states, while if $k=2m+1$ is odd, then the output superposition is also odd. 
Analytical expressions for the $CV$ states $\ket{\Psi_{k}^{(0)}(y_1)}$ and$\ket{\Psi_{k}^{(2)}(y_1,B)}$, as well as their normalization coefficients, are presented in the supplementary material. 
The initial squeezing parameter $y$ of the reference $SMSV$ state decreases by $1+B$ times, i.e., it becomes equal $y_1=y⁄(1+B)$, where introduced $BS$ parameter$ B=(1-t^2 )⁄t^2 $ \cite{30}\cite{32} allows us to express both transmission $T=t^2=1⁄(1+B)$ and reflection $R=r^2=1-t^2=B⁄(1+B)$ coefficients through it.
As for the normalization factors, they are determined by the analytical function $Z(y_1)=1⁄\sqrt{1-4y_1^2}$. 
So the normalization factors of the $CV$ states $\ket{\Psi_{k}^{(0)}(y_1)}$ are the derivatives of the function $Z(y_1)$, i.e. $Z^{(2m)}(y_1)= dZ^{2m}⁄dy_1^{2m} $ for $k=2m$ and $Z^{(2m+1)}(y_1)=dZ^{(2m+1)}⁄dy_1^{(2m+1)}$ for odd $CV$ stares. The normalization factors $G_k^{(2)}(y_1,B)$ of the $CV$ states $\ket{\Psi_{k}^{(2)}(y_1,B)}$ are more complex. 
They are polynomials with derivatives of the function $Z(y_1)$, which are presented in supplementary material.
\par The $CV$ states with the same subscript $k$ but with different superscripts are not orthogonal, i.e., their overlap ${\bra{\Psi_{k}^{(0)}(y_1,B)}\ket{\Psi_{k}^{(2)}(y_1,B)}}\neq 0$, partly because they have the same parity. Therefore, the normalization factor of the output $CV$ state \eqref{eq:3} involves additional term proportional to $\cos \varphi$

\begin{equation}\label{eq:5}
    G_k^{(0,2)}(y_1,B,\varphi) = 1 + b_2^{2}b_k^{(2)2} + \frac{2b_2b_k^{(2)}J_k^{(02)}\cos{\varphi}}{\sqrt{Z^{(k)}(y_1)G_k^{(2)}(y_1,B)}}
\end{equation}

\!\!\!\!\!\!where the cross term $ J_k^{(02)}=\sqrt{(Z^{(k)}(y_1)G_k^{(2)}(y_1,B)} {\bra{\Psi_{k}^{(0)}(y_1,B)}\ket{\Psi_{k}^{(2)}(y_1,B)}}$ is presented in the supplementary material. Thus, using quantum engineering with two $SMSV$ states, which involves measuring a certain number of photons in the auxiliary measurement mode in \hyperref[Figure.1]{Figure.~\ref*{Figure.1}}, it is possible to realize the probe $CV$ state in which the parameter $\varphi$ is already encoded. Note that in the input auxiliary state \eqref{eq:2} this interference term is absent since $\bra{0}\ket{2}=0$, and when directly measuring state no change in the average photon number is observed with changing $\varphi$. Using the expressions for the amplitudes of the hybrid entangled state from the supplementary material, one can derive the exact normalized distribution of generation of the measurement induced probe $CV$ states of definite parity \eqref{eq:3} parameterized by parameter $\varphi$

\begin{equation}\label{eq:6}
    P_k^{(0,2)}(\varphi) = \frac{\sqrt{1-4y_1^2(1+B)^2}(y_1B)^kZ^{(k)}(y_1)}{N_2k!}G_k^{(02)}(y_1,B,\varphi)
\end{equation}

In classical interference, an electromagnetic wave can travel in a superposition of two paths and interferes with itself either constructively or destructively depending on the relative phase between the two paths. Due to the incomplete distinguishability of the $CV$ components of the measurement induced state of a certain parity in equation \eqref{eq:3}, the average number of photons $\langle n^{(02)}_{k\varphi}\rangle= \bra{\Psi_{k,\varphi}^{(02)}(y_1,B)}n\ket{\Psi_{k,\varphi}^{(02)}(y_1,B)}$ can also oscillate with a change in phase $\varphi$ as

\begin{equation}\label{eq:7}
    \langle n^{(02)}_{k\varphi}\rangle= \frac{\langle n^{(0)}_{k}\rangle + b_2^{2}b_k^{(2)2}\langle n^{(2)}_{k}\rangle}{G_k^{(02)}(y_1,B,\varphi)}(1 + V_k^{(02)}\cos \varphi)
\end{equation}

\!\!\!\!\!\!where the mean number of photons in $CV$ components $\ket{\Psi_k^{(0)}(y_1,B)}$ and $\ket{\Psi_k^{(2)}(y_1,B)}$, are defined through their normalization factors by

\begin{equation}\label{eq:8}
    \langle n^{(0)}_{k}\rangle= y_1\frac{Z^{(k+1)}(y_1)}{Z^{(k)}} , \,\,\,\,\,\,\,\,\,\,\,\,\,
    \langle n^{(2)}_{k}\rangle= \frac{y_1\frac{d}{dy_1}G^{(2)}_k(y_1,B)}{Z^{(k)}}
\end{equation}

The non-zero term 

\begin{equation}\label{eq:9}
    \langle n^{(02)}_{k}\rangle=  
    \bra{\Psi_{k}^{(0)}(y_1,B)}n\ket{\Psi_{k}^{(2)}(y_1,B)} = 
    \frac{y_1\frac{d}{dy_1}J_k^{(02)}}{\sqrt{Z^{(k)}Z^{(k)}}}
\end{equation}

arising from the incomplete distinguishability of the $CV$ components is responsible for the interference term due to which the average number of photons can change periodically with the change of phase $\varphi$. By analogy with classical interference, the term

\begin{equation}\label{eq:10}
    V^{(02)}_{k\varphi} = \frac{2b_{2}b_k^{(2)}\langle n^{(02)}_{k}\rangle}{\langle n^{(0)}_{k}\rangle + b_2^{2}b_k^{(2)2}\langle n^{(2)}_{k}\rangle}
\end{equation}

\!\!\!\!\!\!could be recognized for its visibility. However, there is a difference from the standard interference pattern. The first term in \eqref{eq:7} before the bracket contains a factor $G_k^{(02)}(y_1,B,\varphi)$  in the denominator, which also depends on the estimated phase $\varphi$, which can complicate the shape of the interference curve, in contrast to the situation when the normalization factor $G_k^{(02)}(y_1,B,\varphi)$ would not contain a factor proportional to the $\cos \varphi$. 
\par One of the remarkable features of the approach is that the output signal interference measured directly, as shown in \autoref{Figure.1}, is determined by the phase obtained from the auxiliary state \eqref{eq:2}. The relative phase $\varphi$ can be derived from the output signal interference pattern. This distinguishes this technique from conventional $MZ$ interferometry \cite{18}, where the phase dependent interference pattern arises due to the superposition of light propagating along two arms. Moreover, the corresponding interference curve \eqref{eq:7} can be obtained in the case of arbitrary $k$-photon subtraction and for any values of $S$ and $B$, which can only enhance the significance of the proposed method. The minimum possible error $\triangle \varphi$ in estimating the unknown phase shift $\varphi$ is determined by the quantum Cramer-Rao bound $\triangle\varphi_{qcr\,k}^{(02)} = 1 / \sqrt{F_k^{(02)}}\leq \ket{\partial _{\varphi}\Psi_k^{(02)}} = \partial\ket{\Psi_k^{(02)}}/\partial{\varphi}$, which is derived through the quantum Fisher information of the output state in equation \autoref{eq:3}: $F_k^{(02)} = 4 \Bigg ({\bra{\partial_{\varphi}\Psi_{k}^{(02)}}\ket{\partial_{\varphi}\Psi_{k}^{(02)}}}-\lvert{\bra{\partial_{\varphi}\Psi_{k}^{(02)}}\ket{\Psi_{k}^{(02)}}}\rvert^2\Bigg)$, 
where $\ket{\partial _{\varphi}\Psi_k^{(02)}} = \partial\ket{\Psi_k^{(02)}}/\partial{\varphi}$ means the derivative of the output state with respect to the parameter $\varphi$. It is given by

\begin{equation}\label{eq:11}
    F^{(02)}_{k}(y_1,B,\varphi) = 4 X^{(02)2}_k\Bigg(R_k^{(02)2}\Bigg(1 - \frac{b_2^2b_k^{(2)2}}{G_k^{(02)}}\Bigg) - 2\frac{b_2b_k^{(2)}R_k^{(02)}}{\sqrt{G_k^{(02)}}}\cos\varphi-1\Bigg)
\end{equation}

where new quantities are introduced

\begin{equation}\label{eq:12}
    R^{(02)}_{k} = \frac{\sqrt{Z^{(k)}(y_1)G_k^{(2)}(y_1,B)G_k^{(02)}(y_1,B,\varphi)}}{J_k^{(02)}(y_1,B)}
\end{equation}

\begin{equation}\label{eq:13}
    X^{(02)}_{k} = \frac{b_2b_k^{(2)}J_k^{(02)}(y_1,B)}{G_k^{(02)}(y_1,B,\varphi)\sqrt{Z^{(k)}(y_1)G_k^{(2)}(y_1,B)}} = \frac{b_2b_k^{(2)}}{R_k^{(2)}\sqrt{G_k^{(02)}(y_1,B,\varphi)}}
\end{equation}

As follows from expression \autoref{eq:11} the $QFI$ is completely determined by the already introduced normalization factors, the parameters $b_{2}$, $b_k^{(2)}$ and the cross term $J_{k}^{(02)}$. The $QFI$ can change periodically with the change of $\varphi$, but, in general, it strongly depends on three input continuous parameters $S$,$B$ and $b_2$ (or parameter $S_2\ll1$) and one discrete parameter $k$ which can significantly change the shape of the $QFI$.

\begin{figure}[htbp]
\centering\includegraphics[width=0.75\textwidth]{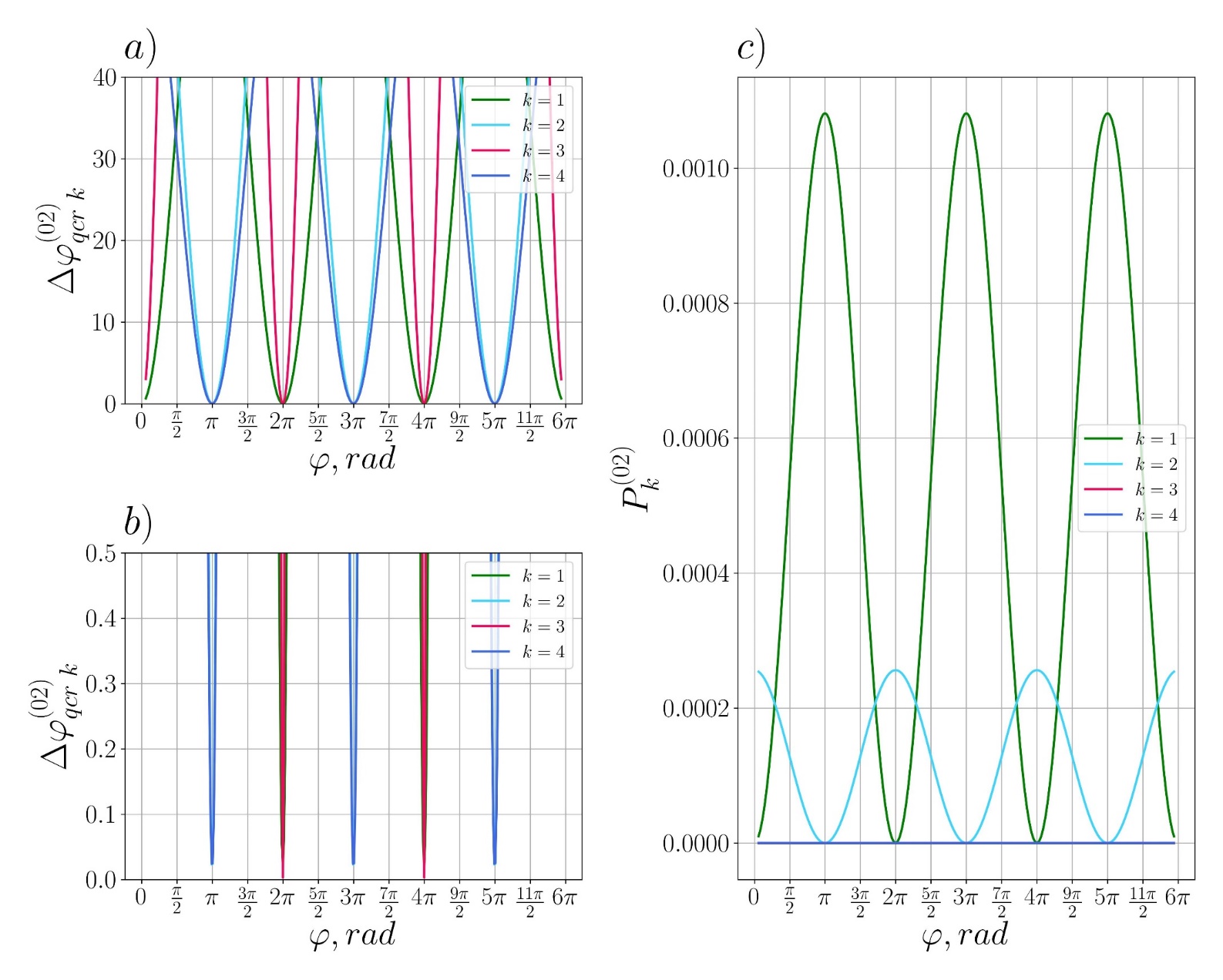}
\caption{(a-c) Periodic dependences of the $QCR$ boundary $\triangle\varphi_{qcr\,k}^{(02)}$ on the value of the estimated phase $\varphi$ for different values of k with $S_2=0.3 \,\,\,dB$ for any $k=1,\,2,\,3,\,4$ and for such $S$ and B (the $S$ and $B$ are different for different k), which provide the minimum value of the utmost bound at the corresponding points $\varphi_1^{(even)}=\pi,\,\varphi_3^{(even)}=3\pi,\,\varphi_5^{(even)}=5\pi$ and $ \varphi_2^{(odd)}=2\pi,\varphi_4^{(odd)}=4\pi$. The graphs in (a) represent the general form of the dependencies, and in (b) the same curves, but on a reduced scale along the vertical axis. In (b) the curves are grouped by parity, so when scaled down along the vertical axis they may appear as one curve rather than two. The corresponding periodic dependencies of the probabilities $P_k^{(02)}$ to generate the $CV$ states \autoref{eq:3} on the estimated parameter $\varphi$ are presented in (c). The curves $P_3^{(02)}$ and $P_4^{(02)}$ are almost equal to each other and therefore appear as a single horizontal line at the chosen scale along the vertical axis.}
\label{Figure.2}
\end{figure}

\!\!\!\!\!\!\!\! \hyperref[Figure.2]{Fig.~\ref*{Figure.2}(a)}
  shows the dependence of the $QFI$ boundary $\triangle\varphi_{qcr\,k}^{(02)}$ on the parameter $\varphi$ at $S_2=0.3\,\,\, dB$ and $k=1,\,2,\,3,\,4$. These curves are constructed by selecting such values of $S$ and $B$ from certain regions that they ensure the minimum value of $\triangle\varphi_{qcr\,k}^{(02)}$ at the some point $\varphi$. In general, the periodic curves can oscillate with a fairly large amplitude as $\varphi$ changes. Those values of $\varphi$ that demonstrate sufficiently large values of $\triangle\varphi_{qcr\,k}^{(02)}>1$ may not be of practical interest. However, there are values of $\varphi$ and regions around them that can provide the $QCR$ boundary with rather small values $\triangle\varphi_{qcr\,k}^{(02)}\ll1$ which is of practical interest. In total, five such rather narrow downward peaks are shown in \hyperref[Figure.2]{Fig.~\ref*{Figure.2}(b)}: three for even and two for odd $k$. 
We are going to denote the values as $\varphi_1^{(even)}=\pi,\varphi_3^{(even)}=3\pi,\varphi_5^{(even)}=5\pi$ and $ \varphi_2^{(odd)}=2\pi,\varphi_4^{(odd)}=4\pi$. The minimum of the $QCR$ limit is observed at these points, and the choice of $S$ and $B$ is due to the fact that the values of $\triangle\varphi_{qcr\,k}^{(02)}$ take minimum values at the points. As follows from \hyperref[Figure.2]{Figure.~\ref*{Figure.2}(b)}, the oscillation period of $QCR$ boundary for both even and odd $CV$ state \autoref{eq:3}  is $2\pi$. The downward-pointing peaks for the even and odd $CV$ states are grouped together. Due to the rather close intersection of even $\triangle\varphi_{qcr\,2}^{(02)}$, $\triangle\varphi_{qcr\,4}^{(02)}$ and odd dependencies $\triangle\varphi_{qcr\,1}^{(02)}$,$\triangle\varphi_{qcr\,3}^{(02)}$ in the \hyperref[Figure.2]{Figure.~\ref*{Figure.2}(b)}, they look like one curve near the corresponding values $\varphi_{i}^{(even)}$,$\varphi_{j}^{(odd)}$ with $i$$=1,3,5$ and $j$$=2,4$. \hyperref[Figure.2]{Figure.~\ref*{Figure.2}(c)} shows the dependencies of the probabilities $P_k^{(02)}$ of generating the $CV$ states of a certain parity in equation \eqref{eq:3} on the estimated parameter $\varphi$ for the same values of $S$, $B$ and $S_2$ as in the construction of \hyperref[Figure.2]{Figure.~\ref*{Figure.2}}.
The probabilities also fluctuate with change in $\varphi$ with a period $2\pi$. They are in phase with $QCR$ boundary $\triangle\varphi_{qcr\,k}^{(02)}$. 
The maximum and minimum values of probabilities are also observed at points $\varphi_{i}^{(even)}$,$\varphi_{j}^{(odd)}$. The probabilities acquire fairly large values (less than 0.2 percent), and further increase in the probabilities is limited by small values of $S$ and $S_2$, due to which the generation of the measurement induced $CV$ state with $k=0$ dominates over all other probabilities.

\begin{figure}[htbp]
\centering\includegraphics[width=0.75\textwidth]{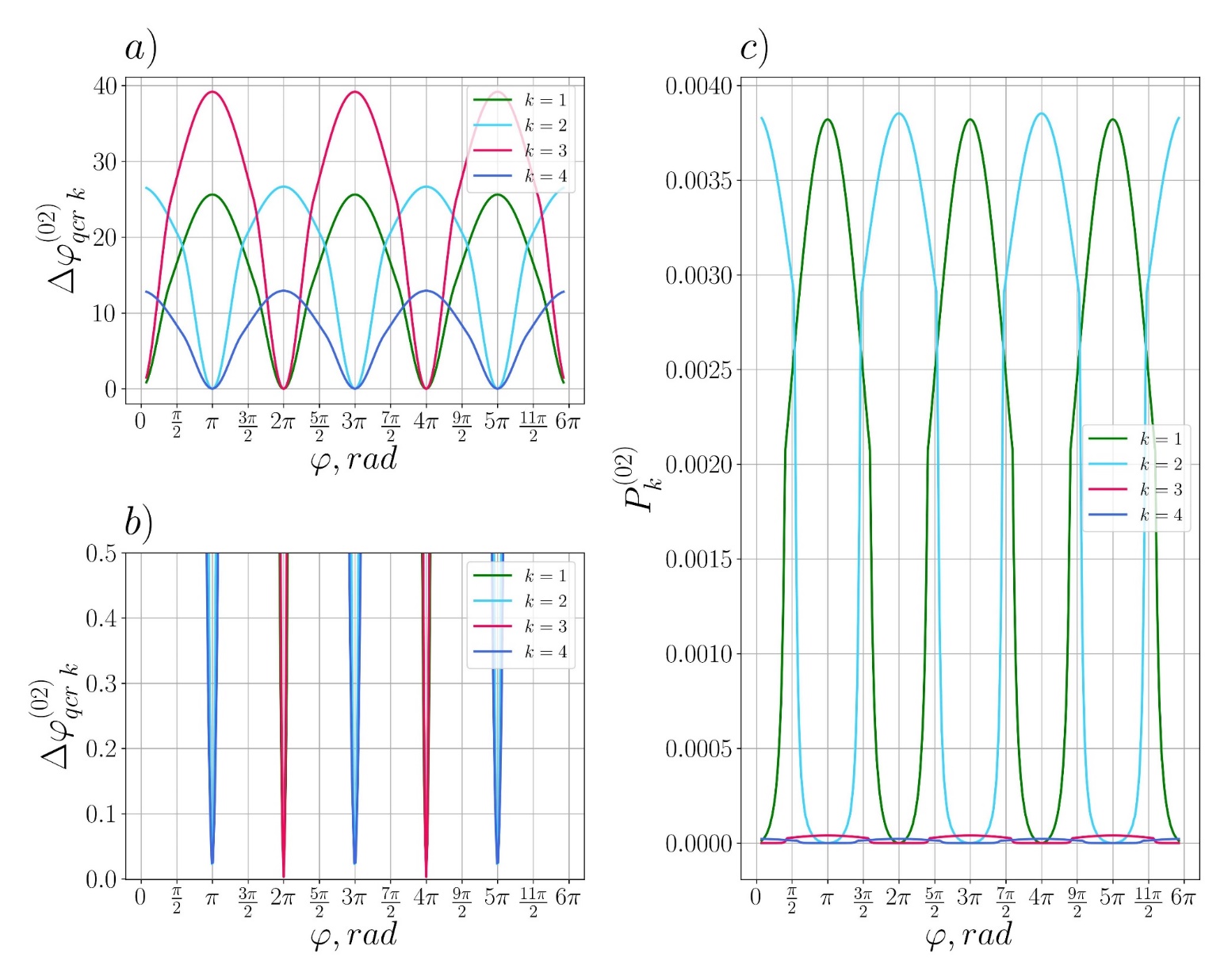}
\caption{(a-c). Optimized periodic dependencies of the $QCR$ boundary $\triangle\varphi_{qcr\,k}^{(02)}$ on $\varphi$ for $S_2=0.3\,\,\,dB$ on (a) large and (b) small scale along the vertical axis. Optimization is performed based on parameters $S$ and $B$. A periodic dependence is observed with minimum values at the same points as in \hyperref[Figure.2]{Figure.~\ref*{Figure.2}}. Optimization allows to reduce the amplitude of $QCR$ boundary oscillations. Periodic dependencies of the probabilities $P_k^{(02)}$ on $\varphi$, optimized for $S$ and $B$, are shown in (c).}
\label{Figure.3}
\end{figure}

If we perform optimization (finding the minimum value of $QCR$ boundary) for the parameters $S$ and $B$ at each point $\varphi$, then the dependencies of $\triangle\varphi_{qcr\,k}^{(02)}$ on $\varphi$ are shown in \hyperref[Figure.3]{Figure.~\ref*{Figure.3}(a)}. Optimization is carried out in certain ranges of change of initial parameters  $S$ and $B$, while $S_2$ remains the same, i.e. $S_2=0.3\,\,dB$ for all $k=1,2,3,4$. Oscillations in the $QCR$ boundary are also observed when $\varphi$ changes with a period $2\pi$, but with a smaller amplitude compared to the oscillation amplitudes in \hyperref[Figure.2]{Figure.~\ref*{Figure.2}(a)}. There are also regions where $\triangle\varphi_{qcr\,k}^{(02)}\ll1$, which are clearly visible in the reduced scale in \hyperref[Figure.3]{Figure.~\ref*{Figure.2}(b)} as downward-pointing peaks. A grouping of the peaks by their parity is observed, when peaks with $k$ of the same parity almost coincide with each other, which appears as a single peak in \hyperref[Figure.3]{Figure.~\ref*{Figure.3}(b)}. As in the case considered in \hyperref[Figure.2]{Figure.~\ref*{Figure.2}(a,b)}, the $QCR$ boundary $\triangle\varphi_{qcr\,k}^{(02)}$ takes on the minimum values at the same five values of $\varphi_{i}^{(even)}$,$\varphi_{j}^{(odd)}$ $\triangle\varphi_{qcr\,k}^{(02)}$. Optimization also makes it possible to increase the probability $P_k^{(02)}$ of measurement results in the measurement mode as shown in \hyperref[Figure.3]{Figure.~\ref*{Figure.3}(c)}. The probabilities realized for those $S$ and $B$ that optimize the $QCR$ boundary oscillate with a period $2\pi$ as the parameter $\varphi$ changes as in \hyperref[Figure.2]{Figure.~\ref*{Figure.2}(c)}.  
Using the photon number operator as a detection observable in \hyperref[Figure.1]{Figure.~\ref*{Figure.1}} is a natural as it does not require post-data processing, and the approach only demands an error estimate. The phase uncertainty can be obtained from the error propagation computation

\begin{equation}\label{eq:14}
    \triangle\varphi_{k}^{(02)} = \frac{\triangle n^{(02)}_k}{\vert \partial\langle n^{(02)}_{k,\,\varphi}\rangle\vert}
\end{equation}

where $\triangle n^{(02)}_k$ = $\sqrt{\langle n^{(02)2}_{k,\,\varphi}\rangle-\langle n^{(02)}_{k,\,\varphi}\rangle^2}$ is the standard deviation of the number of photons with second moment of the photon number operator

\begin{equation}\label{eq:15}
    \langle n^{(02)2}_{k,\,\varphi}\rangle = \frac{1}{G_k^{(02)}}\Bigg(
    \frac{\bigl(y_1\frac{d}{dy_1}\bigl)^2Z^{(k)}(y_1)}{Z^{(k)}(y_1)} + 
    b_2^2b_k^{(2)2}\frac{\bigl(y_1\frac{d}{dy_1}\bigl)^2G^{(2)}_k(y_1)}{G^{(2)}_k(y_1)} + 
    2b_2b_k^{(2)}\frac{\bigl(y_1\frac{d}{dy_1}\bigl)^2J_k^{(02)}}{J_k^{(02)}}\cos \varphi \Bigg)
\end{equation}

and the derivative of the average number of photons with respect to the parameter $\varphi$ is determined by the expression

\begin{equation}\label{eq:16}
    \partial{\langle n^{(02)}_{k\,\varphi}\rangle} = \frac{d\langle n^{(02)}_{k\,\varphi}\rangle}{d \varphi} = 
    2X_k^{(02)}\sin \varphi \Bigg(\langle n^{(02)2}_{k\,\varphi}\rangle - \frac{y_1\frac{d}{dy_1}J_k^{(02)}}{J_k^{(02)}} \Bigg). 
\end{equation}

\begin{figure}[htbp]
\centering\includegraphics[width=0.75\textwidth]{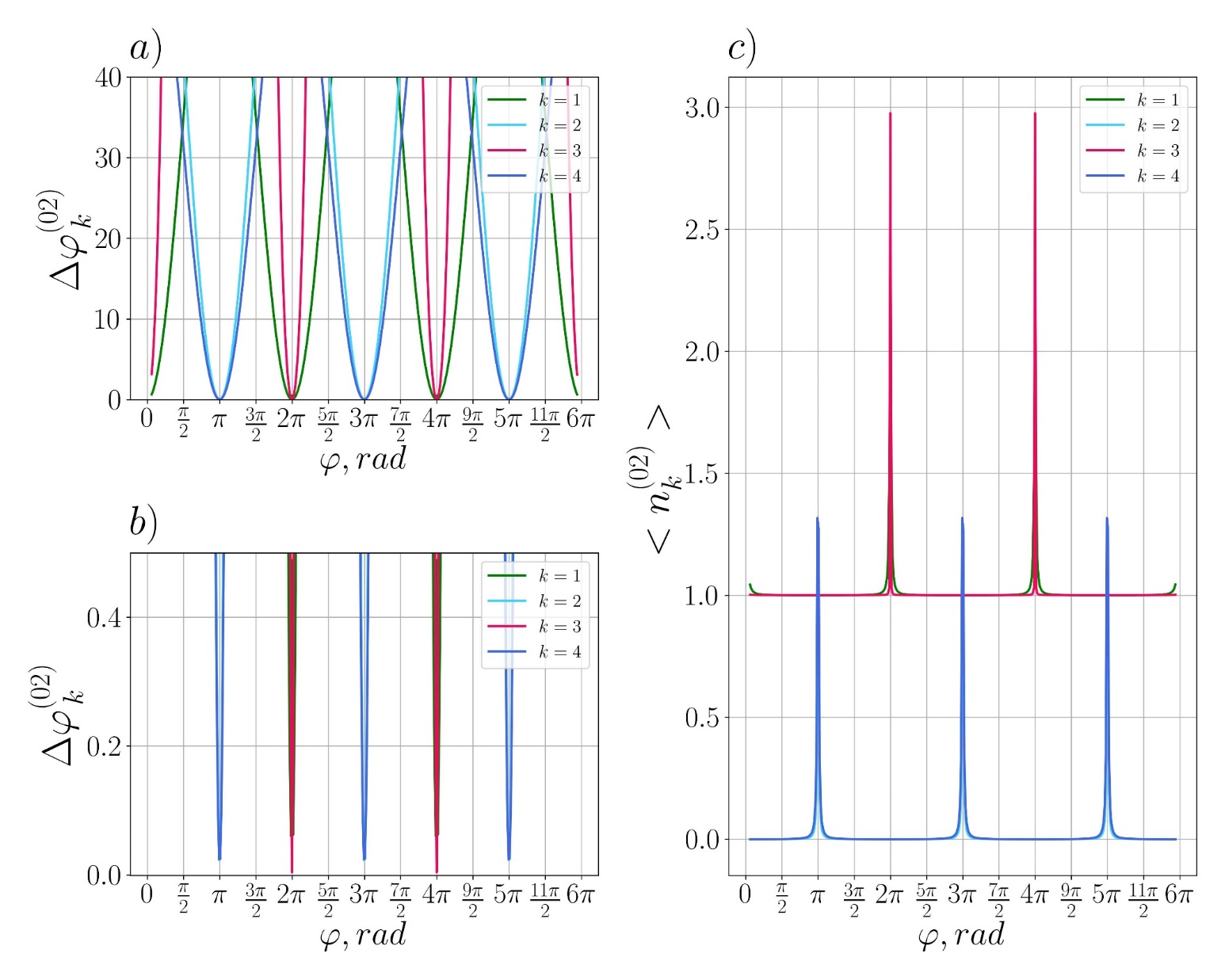}
\caption{(a-c). (a) Periodic dependences of the phase uncertainty $\triangle\varphi_{k}^{(02)}$ in equation (14), obtained by direct measurement of the number of photons, on the parameter $\varphi$ for the same values of the parameters $S$, $B$ and $S_2$ that are used in constructing \hyperref[Figure.2]{Figure.~\ref*{Figure.2}}. The minimum values of $\triangle\varphi_{k}^{(02)}$ on a reduced scale in (b) are at the same values of $\varphi$ as the $QCR$ boundary $\triangle\varphi_{qcr\,k}^{(02)}$, so that $\triangle\varphi_{k}^{(02)}-\triangle\varphi_{qcr\,k}^{(02)}$$\ll1$, which may indicate saturation of such a measurement. The downward-pointing peaks of $\triangle\varphi_{k}^{(02)}$ in (b) are anticorrelated with the upward-pointing peaks of the average number of photons $\langle n^{(02)}_{k}\rangle$ in $CV$ states in (c) at the same values of $\varphi_1^{(even)}=\pi$,$\varphi_3^{(even)}=3\pi$,$\varphi_5^{(even)}=5\pi$ and $\varphi_2^{(odd)}=2\pi$,$\varphi_4^{(odd)}=4\pi$. Due to the grouping of even and odd dependencies, they may appear as one curve instead of two for each $\varphi_i^{(even)}$ and $\varphi_j^{(odd)}$ in (b) and (c).}
\label{Figure.4}
\end{figure}

\hyperref[Figure.4]{Figure.~\ref*{Figure.4}(a)} shows the dependence of $\triangle\varphi_{k}^{(02)}$ on the variable $\varphi$ for those values $S$,$S_2$ and B for which the \hyperref[Figure.2]{Figure.~\ref*{Figure.2}(a-c)} are constructed, that is, $S_2=0.3\,\,dB$ is the same for all graphs and the values of $S$ and $B$ depend on $k$. So, for $k=1$ we have taken $S=0.305 \,\,dB$ and $B=3.527$ as when constructing the graphs in \hyperref[Figure.2]{Figure.~\ref*{Figure.2}(a-c)}. This dependence can also be characterized as periodic with a period $2\pi$. Since we are interested in the estimation errors that take the smallest values $\triangle\varphi_{k}^{(02)}\ll1$, in \hyperref[Figure.4]{Fig.~\ref*{Figure.4}(b)} we show the behavior of these dependencies on a smaller scale in order to highlight those regions in which this condition is satisfied. In general, such a behavior of the phase uncertainty is related to how the average number of photons $\langle n^{(02)}_{k\,\varphi}\rangle$ in the measurement induced $CV$ state changes with varying $\varphi$, as shown in \hyperref[Figure.4]{Figure.~\ref*{Figure.4}(c)}. Apart from the parity grouping (even $\langle n^{(02)}_{2m\,\varphi}\rangle$ and odd $\langle n^{(02)}_{2m+1\,\varphi}\rangle$, the average number of  photons can vary little over most of the range of variations $\varphi$, with except for small regions around which sharp jumps in $\langle n^{(02)}_{k\,\varphi}\rangle$ are observed. The sharp amplification ranges in $\langle n^{(02)}_{2m\,\varphi}\rangle$ are related to the already mentioned values $\varphi_{i}^{(even)}$,$\varphi_{j}^{(odd)}$ with $i=1,3,5$ and $j=2,4$, which also determine the minimum values of $\triangle\varphi_{k}^{(02)}$.
Comparing the behavior of $\langle n^{(02)}_{k\,\varphi}\rangle$ and $\triangle\varphi_{k}^{(02)}$, one can notice a correlation between a sharp increase in the average number of photons and sharp drop in the estimation error, which is especially noticeable at points $\varphi_i^{(even)}$,$\varphi_j^{(odd)}$. There is also a difference in the initial values from which the sharp increase in the average number of photons starts as $\varphi$ approaches either $\varphi_i^{(even)}$ or $\varphi_j^{(odd)}$. This occurs because the even $CV$ probe states approach the vacuum state, while the odd $CV$ states approach single photon in the case of $S\rightarrow0$.
\par If we compare the minimum values of the estimation error $\triangle\varphi_{k}^{(02)}$, obtained using the error propagation method, with the minimum possible errors associated with $\triangle\varphi_{qcr\,k}^{(02)}$ at the points $\varphi_i^{(even)}$ or $\varphi_j^{(odd)}$, we can notice their close similarity in magnitude. The data for comparison are collected in
\hyperref[Table.1]{Table.\ref{Table.1}}
 for $k=1,\,2,\,3,\,4.$ When comparing, the parameter values used to construct \hyperref[Figure.2]{Figure.~\ref*{Figure.2}}  and \hyperref[Figure.4]{Figure.~\ref*{Figure.4}} are selected, that is, $S_2=0.3\,\,dB$ and corresponding values of $S$ and $B$ found numerically. As can be seen from the \hyperref[Table.1]{table}, the condition $\triangle\varphi_{k}^{(02)}$>$\triangle\varphi_{qcr\,k}^{(02)}$  is fulfilled for all cases $ k=1,\,2,\,3,\,4$ considered, but the phase uncertainty $\triangle\varphi_{k}^{(02)}$ can take values very close to the $\triangle\varphi_{qcr\,k}^{(02)}$, i.e. $\triangle\varphi_{,k}^{(02)}\approx\triangle\varphi_{qcr\,k}^{(02)}$. This is especially noticeable for  $k=2$ and $k=4$, where the difference between $\triangle\varphi_{k}^{(02)}$ and the $QCR$ boundary $\triangle\varphi_{k}^{(02)}$-$\triangle\varphi_{qcr\,k}^{(02)}$ occurs at the fifth decimal place for $k=2$ and at the seventh decimal place for $k=4$. The difference between the two quantities becomes somewhat larger in the case of $k=1$ and $k=3$. However, in the general case, the direct measurement of the average number of photons of the probe  $CV$ states  \autoref{eq:3}, in which information about an unknown parameter $\varphi$ is encoded in the quantum engineering process, when calculated in the vicinity of points $\varphi_i^{(even)}$, $\varphi_j^{(odd)}$, can be considered saturating, i.e. the measurement error calculated by the error propagation method is comparable to the $QCR$ limit. In addition, \autoref{Table.1} also presents the values of the average number of photons at the points, i.e. $\langle n^{(02)}_{2}\rangle$=$\langle n^{(02)}_{2\,\varphi = \pi}\rangle$, $\langle n^{(02)}_{4}\rangle$=$\langle n^{(02)}_{2\,\varphi = \pi}\rangle$ for even $CV$ states and $\langle n^{(02)}_{1}\rangle$=$\langle n^{(02)}_{1\,\varphi = 2\pi}\rangle$, $\langle n^{(02)}_{3}\rangle$=$\langle n^{(02)}_{3\,\varphi = 2\pi}\rangle$ for odd $CV$ states, respectively.

\begin{table}[htbp]
\centering
\begin{tabularx}{\textwidth}{|l|X|X|X|}
\hline
$k$ & $\triangle\varphi_{qcr\,\,k}^{(02)}(min)$ & $\triangle\varphi_{k}^{(02)}(min)$ & $\langle n^{(02)}_{k}\rangle(max)$ \\
\hline
$1$ & $0.01888959433438779$ & $0.06030172518672809$ & $2.512252277788733$ \\
$2$ & $0.02300605930321999$ & $0.023018208739094614$ & $1.0177369871306052$\\
$3$ & $0.0030162367758169123$ & $0.0038579400567110935$ & $2.9757740087801583$ \\
$4$ & $0.024241076566853478$ & $0.024241298132351628$ & $1.3170327469211816$\\
\hline
\end{tabularx}
\caption{Comparison of the $QCR$ boundary $\triangle\varphi_{qcr\,\,k}^{(02)}$ in equation \autoref{eq:11} and the minimum error $\triangle\varphi_{k}^{(02)}$ in equation \autoref{eq:14} resulting from the error propagation formula, when the average number of photons in the measurement induced $CV$ state of a certain parity \autoref{eq:3} is directly measured at points $\varphi_{1}^{(even)} = \pi$ for $k=2,\,4$ and $\varphi_{2}^{(odd)} = 2\pi$ for $k=1,\,3$. The last column shows the maximum values $\langle n^{(02)}_{k}\rangle$ of the average number of photons of the probe $CV$ state parameterized by the parameter $\varphi$ at the points $\varphi_{1}^{(even)}$ and $\varphi_{1}^{(odd)}$, respectively. The minimum values of $\triangle\varphi_{k}^{(02)}$ and $\triangle\varphi_{qcr\,\,k}^{(02)}$ and maximum values of $\langle n^{(02)}_{k}\rangle$ are calculated for the same values of S,$\,\,S_2$ and $B$ for each $k$.}
\label{Table.1} 
\end{table}

\begin{figure}[htbp]
\centering\includegraphics[width=0.75\textwidth]{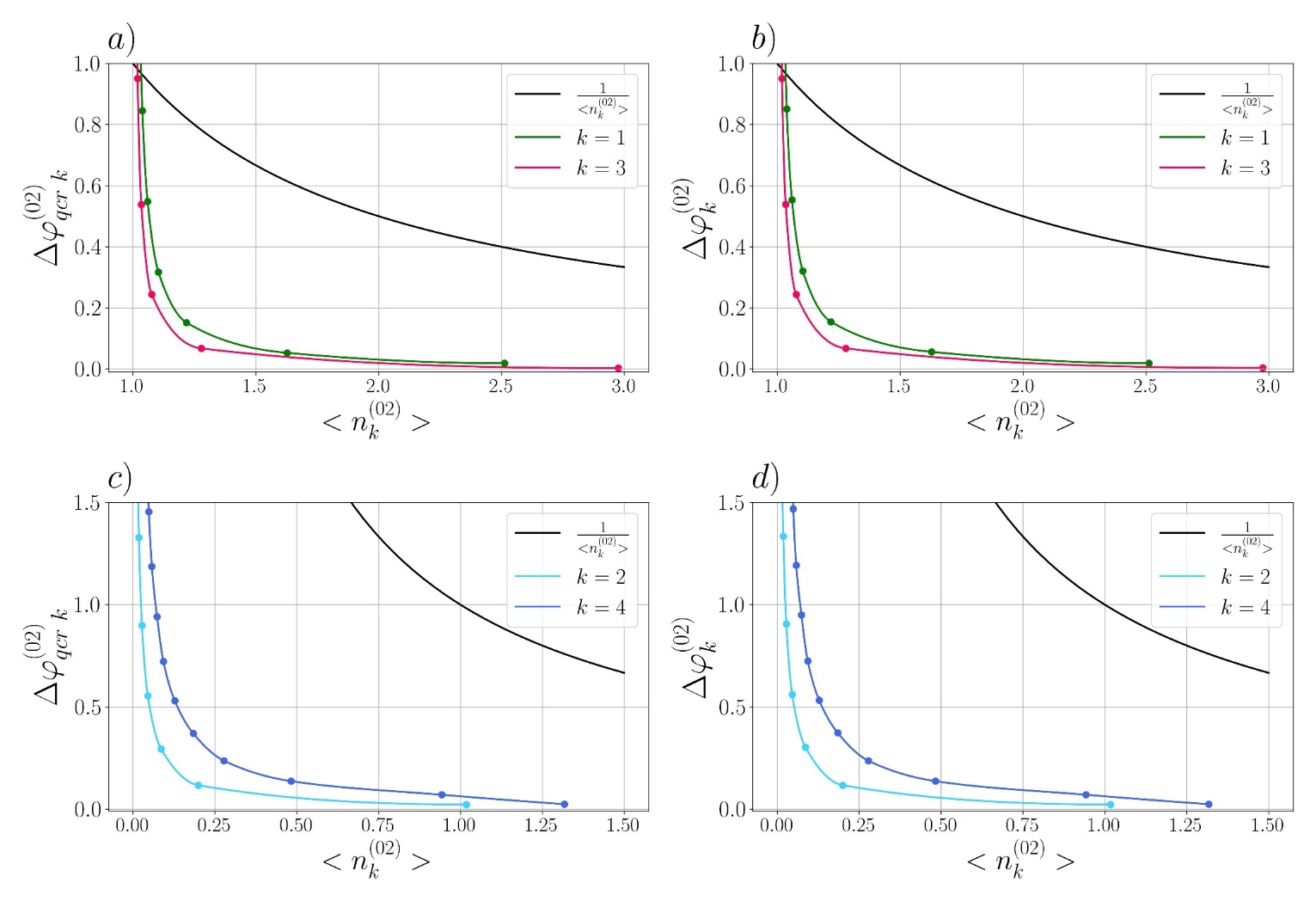}
\caption{(a-d). Dependence of (a) the $QCR$ boundary $\triangle\varphi_{qcr\,k}^{(02)}$ and (b) the phase uncertainty $\triangle\varphi_{k}^{(02)}$ on the average number of photons $\langle n^{(02)}_{k}\rangle$ in odd $CV$ states \autoref{eq:3}  with $k=1,\,3$ for phase shift $\varphi=2\pi-\delta$ with $\delta \rightarrow 0$. Similar dependencies of (c) the QCR boundary $\triangle\varphi_{qcr\,k}^{(02)}$ and (d) the phase uncertainty $\triangle\varphi_{k}^{(02)}$ on the average number of photons $\langle n^{(02)}_{k}\rangle$ but for even $CV$ states at $k=2,\,4$ near the phase shift $\varphi_1^{(even)}=\pi$. All four figures also show the dependence of $\langle n^{(02)}_{k}\rangle^{-1}$, which is significantly higher than the corresponding dependences $\triangle\varphi_{qcr\,k}^{(02)}$ and $\triangle\varphi_{k}^{(02)}$, indicating the possibility of achieving sub-Heisenberg precision using the probe state \autoref{eq:3}  encoding information about $\varphi$. The average number of photons $\langle n^{(02)}_{k}\rangle$ of the probe $CV$ state with encoded information $\varphi$ follows from expression \eqref{eq:7} for (a,b) $\varphi=2\pi-\delta$ and (c,d) $\varphi=\pi-\delta$ with $\delta\rightarrow 0$.  }
\label{Figure.5}
\end{figure}

As can be seen from the graphs in \hyperref[Figure.4]{Figure.~\ref*{Figure.4}}, a sharp increase in the average number of photons near the values $\varphi_{i}^{(even)}$,$\varphi_{j}^{(odd)}$ leads to a sharp increase in its sensitivity. Therefore, it is interesting to trace the behavior of the estimate errors with an increase in the average number of photons in $CV$ states \autoref{eq:3} near these points. By the number of photons we mean the average number of photons$\langle n^{(02)}_{k}\rangle$=$\langle n^{(02)}_{k\,\varphi}\rangle$ in the probe $CV$ state, which already contains information about the encoded value $\varphi$. 
In what follows, we are going to use the values of the parameters $S$, $S_2$ and $B$ that are used to construct the graphs in \hyperref[Figure.3]{Figure.~\ref*{Figure.3}}, that is, those that optimize the $QCR$ boundary $\triangle\varphi_{qcr\,\,k}^{(02)}$ at each point $\varphi$. \hyperref[Figure.5]{Figure.~\ref*{Figure.5}} (a) shows the dependence of the $QCR$ boundary $\triangle\varphi_{qcr\,\,k}^{(02)}$ on the average number of photons $\langle n^{(02)}_{k}\rangle$ in odd $CV$ states with $k=1,\,3$, where both of the parameters are obtained by choosing the values of $\varphi$ in the left neighborhood of $\varphi_2^{(odd)}=2\pi$, i.e. $\varphi=2\pi-\delta$, where $0<\delta \ll1$ with a gradual decrease of $\delta$ to zero $\delta\rightarrow 0$.
\hyperref[Figure.5]{Figure.~\ref*{Figure.5}(b)} demonstrates the dependence of $\triangle\varphi_{k}^{(02)}$ on the average number of photons $\langle n^{(02)}_{k}\rangle$ in odd $CV$ states, obtained using the same approach. In general, the dependencies of $\triangle\varphi_{qcr\,k}^{(02)}$ and $\triangle\varphi_{k}^{(02)}$ for the same $\langle n^{(02)}_{k}\rangle$ look similar to each other in  \hyperref[Figure.5]{Figure.~\ref*{Figure.5}(a,b)}, but the numerical results confirm that there is a small difference between $\triangle\varphi_{qcr\,k}^{(02)}$ and $\triangle\varphi_{k}^{(02)}$ and $\triangle\varphi_{k}^{(02)}$>$\triangle\varphi_{qcr\,k}^{(02)}$ at each point $\langle n^{(02)}_{k}\rangle$. Thus, we confirm that, under certain conditions, increasing the average number of photons leads to a decrease in both the utmost error of the estimate and the practical error, which can be estimated based on the results of processing statistical information. An increase in the sensitivity of the optical scheme in \hyperref[Figure.1]{Figure.~\ref*{Figure.1}} is also observed with increasing number k of subtracted photons, as evidenced by the fact that the curve with $k=3$ lies below the curve with $k=1$. A similar behavior is demonstrated by the even $CV$ states with $k=2,\,4$, for which the dependences of $\triangle\varphi_{qcr\,k}^{(02)}$ and $\triangle\varphi_{k}^{(02)}$ on the average number of photons $\langle n^{(02)}_{k}\rangle$ are shown in  \hyperref[Figure.5]{Figure.~\ref*{Figure.5}(c)} and \hyperref[Figure.5]{(d)}, respectively. These numerical dependencies are obtained in the vicinity of point $\varphi_2^{(even)}=\pi$, that is, for $\varphi=\pi-\delta$ with positive $\delta \rightarrow 0$, and allow us to estimate the rate of decrease of $\triangle\varphi_{qcr\,k}^{(02)}$ and$\triangle\varphi_{k}^{(02)}$ with increasing average number of photons $\langle n^{(02)}_{k}\rangle$. In contrast to the case with odd probe $CV$ states, increasing the number of subtracted photons from $k=2$ to $k=4$ no longer leads to a decrease in either the utmost or realistic errors.

The dependencies in \hyperref[Figure.5]{Figure.~\ref*{Figure.5}}  clearly demonstrate the possibility of increasing the sensitivity of the optical scheme in \hyperref[Figure.1]{Figure.~\ref*{Figure.1}} to the input photons, that is, adding an input photon can only reduce the error in estimating the unknown phase of $\varphi$. An increase in sensitivity per input photon is observed only in the left neighborhood of the points $\varphi_i^{(even)}$,$\varphi_j^{(odd)}$. If we consider the value of $\varphi$ in the right neighborhood of the points $\varphi_i^{(even)}$,$\varphi_j^{(odd)}$, i.e. either $\varphi=\varphi_i^{(even)}+\delta$ or $\varphi=\varphi_j^{(odd)}+\delta$ with $\delta>0$, then in the case the average number of photons begins to decrease, and the quantities $\triangle\varphi_{qcr\,k}^{(02)}$ and $\triangle\varphi_{k}^{(02)}$ start to increase with $\delta$ growing. To compare the rates of sensitivity increase, \hyperref[Figure.5]{Figure.~\ref*{Figure.5}}  also shows the Heisenberg limit graphs $HL=\langle n^{(02)}_{k}\rangle ^{-1}$ \cite{6}\cite{7}\cite{8}, which lie significantly above the already mentioned curves meaning that the conditions $\triangle\varphi_{qcr\,k}^{(02)}$<$\langle n^{(02)}_{k}\rangle ^{-1}$  and $\triangle\varphi_{k}^{(02)}$<$\langle n^{(02)}_{k}\rangle ^{-1}$ are satisfied for all $k$. Thus, quantum engineering of the probe $CV$ state presented in  \hyperref[Figure.1]{Figure.~\ref*{Figure.1}}  makes it possible to implement superior quantum sensor for detecting extremely subtle changes in phase shift $\varphi$ with precision surpassing not only the $SQL$ but also the $HL$. 

\section{Influence of the quantum efficiency of the PNR detector on precision of the estimation}

Above we considered the possibility of using quantum engineering of quantum light generated from probe $SMSV$ state, which can be encoded by an unknown parameter using an auxiliary feebly squeezed state. The above analysis assumes ideal operation of the optical elements, in particular, the use of a $PNR$ detector with unit quantum efficiency. In practice, the quantum efficiency $\eta \neq 1$ of the $PNR$ detector \cite{34}\cite{35} can add additional $CV$ states into the probe parameterized $CV$ state \autoref{eq:3} . 
\par To take this factor into account, it is necessary to use a realistic PNR detector model. The PNR detector with quantum efficiency $\eta$ can be modeled using the positive operator-valued measure $(POVM)$ formalism with $POVM$ elements  $\lbrace\hat{\Pi}_{k},k=0,\ldots,\infty \rbrace$  \cite{33} . Using the general expression for the entangled hybrid state before measurement in auxiliary mode (formula (S2) in supplementary material) and applying the POVM formalism, in the first order in the smallness parameter $1-\eta\ll1$, one obtains expression for output state and its probability

\begin{equation}\label{eq:17}
    \rho_k^{(02)} = \frac{\rho_k^{(02)} + (1-\eta)y_1Bq^{(02)}_{k+1}\rho^{(02)}_{k+1}}{1+(1-\eta)y_1Bq_{k+1\,q}^{(02)}}, 
\end{equation}

\begin{equation}\label{eq:18}
    P_k^{(02)}(\eta) = P_k^{(02)}(\eta=1)\eta^k(1 + (1-\eta)y_1Bq^{(02)}_{k+1\,k}), 
\end{equation}

\!\!\!\!\!\!where parameter associated with the ratio of normalization coefficients $q^{(02)}_{k+1\,k} = \\(Z^{(k+1)}(y_1)G_{k+1}^{(02)})/(Z^{(k)}(y_1)G_{k}^{(02)})$ is introduced and $P_k^{(02)}(\eta=1)$
is given by formula (6). Here the states $\rho_k^{(02)} = \ket{\Psi_k^{(02)}}\bra{\Psi_k^{(02)}}$ and $\rho_{k+1}^{(02)} = \ket{\Psi_{k+1}^{(02)}}\bra{\Psi_{k+1}^{(02)}}$ are formed from $CV$ states with subscripts $k$ and $k+1$ differing from each other by 1, which leads to their mutual orthogonality, since they have different parity. Thus, the output state can also be rewritten as
$\rho_{k}^{(02)} = \lambda_1\rho_{k}^{(02)} + \lambda_2\rho_{k}^{(02)}$ with $\lambda_1 = 1/\bigg(1+(1-\eta)y_1Bq^{(02)}_{(k+1\,k)}\bigg)$ and  $\lambda_2 = (1-\eta)y_1Bq^{(02)}_{k+1\,k}/\bigg(1+(1-\eta)y_1Bq^{(02)}_{k+1\,k}\bigg)$ satisfying the normalization condition $\lambda_1 + \lambda_2 = 1$ This allows us to calculate the $QFI$ of the output state as $F_{k\,\eta}^{(02)} = \lambda_1F_{k}^{(02)} + \lambda_2F_{k+1}^{(02)}$, where $F_k^{(02)}$ and $F_{k+1}^{(02)}$ follow from equation \autoref{eq:11}  with subscripts $k$ and $k+1$, respectively.       
In general, $QFI$ dependencies have qualitatively the same periodic forms as shown in  \hyperref[Figure.2]{Figure.~\ref*{Figure.2}}  and \hyperref[Figure.3]{Figure.~\ref*{Figure.3}} in the case of $1-\eta\ll1$, that is, the maximum values of the $QFI$ are also observed at the corresponding points $\varphi$. The $QFI$ functions with subscripts that differ by one behave in antiphase to each other, that is, if $F_k^{(02)}$ increases, then $F_{k+1}^{(02)}$ decreases and vice versa. This circumstance reduces the value of $QFI$ at the corresponding points $\varphi_i^{(even)}$,\,$\varphi_j^{(odd)}$, which leads to an increase in the values of the $QCR$ boundary compared to the case of using an ideal $PNR$ detector. Overall, this increase in the limiting error is insignificant in the case of $1-\eta\ll1$. For example, we have
$ \triangle\varphi_{qcr\,k\,\eta}^{(02)}=0.051836$ for $k=1$ and $\eta=0.95$ at the point $\varphi_2^{(odd)} = 2\pi$, which, although larger than $\triangle\varphi_{qcr\,k}^{(02)}=0.01889$, but this increase is not critically large, indicating that the intensity measurement can be quite robust to imperfections in the measuring equipment.  

\section{Conclusion}

We have developed a method for the conditional generation of the $CV$ states with encoded phase information to achieve signal measurement precision approaching quantum the Cramer-Rao boundary. In the proposed quantum-enhanced optical interferometer, the probe phase-parameterized $CV$ state of a certain parity is implemented, after which the intensity of the state is measured. Encoding of the original probe state with an unknown phase is achieved by interacting the reference $SMSV$ state with the auxiliary weakly squeezed $SMSV$ state at the beam splitter. The selection of the final nonclassical $CV$ state of definite parity \autoref{eq:3}  from the hybrid entangled state generated by the beam splitter is induced by measuring the exact number of photons in the auxiliary mode. The key difference between the $CV$ state under consideration and the $SMSV$ state lies in its structure. It consists of two $CV$ states of a certain parity, non-orthogonal to each other. A realistic model of the interaction of $SMSV$ states on a beam splitter with arbitrary transmittance and reflectance allows optimizing the output characteristics of the phase-parameterized $CV$ state by varying the squeezing of the initial $SMSV$ states and the $BS$ parameter. In general, the probabilities take on quite large values, with the probability of zero photon subtraction dominating. This is related to the values of the squeezing parameter used in the analysis.
\par The periodical $QCR$ boundary with a period $2\pi$ of the probe $CV$ states of definite parity, parameterized by the phase parameter, takes minimum values at the points $\varphi= \pm\pi k$, where $k=2l+1$ corresponds to even $CV$ states and $k=2l$ to odd $CV$ states. In the vicinity of the points, anticorrelation between the descending peaks on the $QCR$ boundary and the ascending peaks in the average number of photons results in an increase in the sensitivity per input additional photon. As result, the intensity measurement of the measurement induced phase-parameterized $CV$ state of definite parity reaches saturation. Moreover, the $QCR$ boundary is significantly smaller than the reciprocal of the average number of photons in the phase-parametrized probe $CV$ state, indicating the sub-Heisenberg precision of proposed interferometer. 
\par In general, the optical interferometer in  \hyperref[Figure.1]{Figure.~\ref*{Figure.1}}  can operate with both a $PNR$ detector \cite{34}\cite{35} and a second detector measuring the light intensity. Unlike the classical design of a Mach-Zehnder interferometer with passive optical elements \cite{18}  or the SU interferometer with active optical elements \cite{21}  such as optical parametric amplifiers, the optical scheme in  \hyperref[Figure.1]{Figure.~\ref*{Figure.1}}  does not require either a second beam splitter or a second parametric amplifier. The scheme shown in \hyperref[Figure.1]{Figure.~\ref*{Figure.1}} can be particularly useful if the experimenters have a rough estimate of the phase value. In this case, they can add an additional phase shift up to values $\varphi=\pm\pi k$ to improve the estimation precision. This formulation of the problem of quantum engineering of the probe $CV$ state parameterized by an unknown parameter is realistic and feasible in practice, since the $SMSV$ states, which are the basic components of the approach, can be generated deterministically in degenerate optical parametric down conversion. The intensity measurement is quite reliable when creating the probe phase-parameterized $CV$ using $PNR$ detector with non-ideal quantum efficiency. The interferometric scheme in  \hyperref[Figure.1]{Figure.~\ref*{Figure.1}}  also allows for expansion and modification. 

\begin{backmatter}

\bmsection{Funding}
\!\!\!The study was supported by the grant of the Russian Science Foundation No. 25-12-20026, https://rscf.ru/project/25-12-20026/.

\bmsection{Acknowledgment}
 \!\!\!\!\!\!Supplementary material for the manuscript was prepared with the support of the Foundation for the Advancement of Theoretical Physics and Mathematics “BASIS” (Project № 24-1-1-87-1). 

\bmsection{Disclosures}
\!\!\!The authors declare no conflicts of interest.

% \bmsection{Supplemental document}
% \!\!\!See \href{Sup_material.tex}{supplementary material} for supporting content.

\end{backmatter}

\renewcommand{\refname}{5.\,\,\,\,References}

\newpage

\title{Supplementary material for Ultra-precise phase estimation without mode entanglement}\label{chap:Supplementary}

\author{Mikhail S.Podoshvedov,\authormark{1,2} Sergey A. Podoshvedov,\authormark{1,2,*}}

\address{\authormark{1}Laboratory of quantum information processing and quantum computing, South Ural State University (SUSU), Lenin Av. 76, Chelyabinsk, Russia\\
\authormark{2}Laboratory of quantum engineering of light, South Ural State University (SUSU), Lenin Av. 76, Chelyabinsk, Russia}

\email{\authormark{*}sapodo68@gmail.com}

\section*{Supplementary note 1: Superposition of CV states with zero and two photons added}

Here we briefly consider the possibility of forming a target CV state of a certain parity used in estimating an unknown parameter. 
It can be realized by mixing two SMSV states on a beam splitter ($BS$) with arbitrary real transmittance $t>0$ and reflectance $r>0$ satisfying the physical condition $t^2+r^2=1$. The $BS$ transforms the creation operators $a_1^{\dagger}$ and $a_2^{\dagger}$ as $BS_{12}a_1^{\dagger}BS_{12}^{\dagger}=ta_1^{\dagger}-ra_2^{\dagger}$ and $BS_{12}a_2^{\dagger}BS_{12}^{\dagger}=ra_1^{\dagger}+ta_2^{\dagger}$ \cite{chap2:1}. The $SMSV$ state with corresponding notations for squeezing amplitude s and squeezing parameter y are presented in the main text. If one of the $SMSV$ state is the reference, then for the second we use an approximation 

\setcounter{equation}{0}      
\begin{equation}\label{eq:S1}
    \ket{SMSV_{s_2}} \approx \ket{\varphi_2} = 
    \frac{1}{\sqrt{N_2}}(\ket{0} + b_2\exp(i\varphi)\ket{2}), 
    \tag{S1}
\end{equation}

\!\!\!\!\!\!that is relevant in the case of a small squeezing amplitude $s_2\ll1$, 
where the amplitude $b_2$ of the $SMSV$ state, that is,
$b_2 = \tanh s_2 / \sqrt{2} = \sqrt{2}y_2$
in its Fock representation is used and $N_2=1+b_2^2$ is the normalization factor of the additional state.
Mixing the reference $SMSV$ state with the superposition state (\ref{eq:S1}) generates the hybrid entangled state.
Mixing the reference $SMSV$ state with the superposition state (\ref{eq:S1}) generates the hybrid entangled state

\begin{equation}\label{eq:S2}
    BS_{12}\big(\ket{SMSV(y)}_1\ket{\varphi_2}_2\big) = 
    \frac{1}{\sqrt{N_{2}\cosh s}}\sum_{k=0}^{\infty} \left(
        \begin{array}{c}
            c_k^{(0)}(y_1,B)\sqrt{Z^{(k)}(y_1)} \\
            \sqrt{G_k^{(02)}(y_1,B)}\ket{\Psi_k^{(2)}(y_1,B)}_1 \ket{k}_2
        \end{array} 
    \right),
    \tag{S2}
\end{equation}

where the $CV$ states of definite parity are given by

\begin{equation}\label{eq:S3}
    \ket{\Psi_k^{(02)}(y_1,B,\varphi)} = 
    \frac{1}{\sqrt{G_k^{(02)}(y_1,B,\varphi)}}
    \Bigg(\ket{\Psi_k^{(0)}(y_1)} + 
    b_2b_k^{(2)}\exp{(i\varphi)}\ket{\Psi_k^{(2)}(y_1,B)}\Bigg), 
    \tag{S3}
\end{equation}

\!\!\!\!\!\!composed of two $CV$ components $\ket{\Psi_k^{(0)}(y_1)}$ and $\ket{\Psi_k^{(2)}(y_1,B)}$ of the same parity. 
The following notations are used here: the superscript either 0 or 2 indicates the number of input photons, and the subscript k is the number of photons measured (subtracted) in the second measurement mode. These $CV$ states depend on the squeezing parameter $y_1$ reduced by $t^2$ on compared with input value y, i.e. $y_1=yt^2=y⁄(1+B)$, and the beam splitter parameter $B=r^2⁄t^2$ \cite{chap2:1} from which the transmittance and reflectance coefficients of the beam splitter can be expressed as $T=t^2=1⁄(1+B)$ and $R=r^2=B⁄(1+B)$.
The measurement induced $CV$ states of a certain parity \ref{eq:S3} are generated by measuring $k$ photons in the auxiliary (second) mode.
$CV$ states $\ket{\Psi_k^{(0)}}$ and $\ket{\Psi_k^{(2)}}$ are the states of one parity which coincides with the parity of the subtracted (measured in the second mode) photons. If $k=2m$ is an even number, then the measurement induced $CV$ state $\ket{\Psi_{2m}^{(02)}}$ is also even. In case of subtraction of odd number of photons $k=2m+1$, the output $CV$ state $\ket{\Psi_{2m+1}^{(02)}}$ is also odd. The phase factor $\exp(i\varphi)$ is included in each of the possible output superpositions and a new real factor

\begin{equation}\label{eq:S4}
    b_{k}^{(2)}(y_1,B) =\frac{c_k^{(2)}(y_1,B)}{\sqrt{2}c_k^{(0)}(y_1,B)}\sqrt{\frac{G_k^{(2)}(y_1,B)}{Z^{(k)}(y_1)}} =\frac{1}{\sqrt{2}(1 + B)}
    \begin{cases}
            \frac{B}{y_1}\sqrt{\frac{G_0^{(2)}(y_1,B)}{Z(y_1)}}, 
            \,\,\text{if}\,\,\, k = 0 \\[3ex]
			-\frac{2}{y_1}\sqrt{\frac{G_1^{(2)}(y_1,B)}{Z^{(1)}(y_1)}}, & 
            \!\!\!\!\!\!\!\!\!\!\!\!\!\!\!\!\!\!\!\!\!\text{if } k = 1 \\[3ex]
			\frac{k(k-1)}{y_1B}\sqrt{\frac{G_{k}^{(2)}(y_1,B)}{Z^{(k)}(y_1)}}, &\!\!\!\!\!\!\!\!\!\!\!\!\!\!\!\!\!\!\text{if } k > 0 
	\end{cases}
\tag{S4}
\end{equation}

\!\!\!\!\!\!is added to the input $b_2$ so that the amplitude of the $CV$ state $\ket{\Psi_{k}^{(2)}}$ is already a product of initial amplitude factor $b_2$ and measurement-induced one $b_k^{(2)}$, i.e. $b_2$$b_k^{(2)}$. The additional amplitude multiplier $b_k^{(2)}$ is proportional to the ratio of the amplitude $c_k^{(2)}(y_1,B)$ to $c_k^{(0)}(y_1,B)$, which are obtained in the case of using an additional input two photon and vacuum states 

\begin{equation}\label{eq:S5}
    c_k^{(0)}(y_1,B) = (-1)^{(k)}\frac{(y_1B)^{\frac{k}{2}}}{\sqrt{k!}},
\tag{S5}
\end{equation}

\begin{equation}\label{eq:S6}
    c_{k}^{(2)}(y_1,B) =\frac{1}{1 + B}
    \begin{cases} \,\,\,\,\,\,\,\,\,\,\,\,\,\,\,\,\,\,\,\,\,\,\,\,\frac{B}{y_1}, 
            \,\,\,\,\,\,\,\,\text{if}\,\,\, k = 0 \\[3ex]
			\,\,\,\,\,\,\,\,\,\,\,\,\,\,\,\,\,2\sqrt{\frac{B}{y_1}} , \,\,\,\,\,\,\text{if } k = 1 \\[3ex]
			\,\,\,(-1)^k \frac{(y_1B)^{(k/2-1)}}{\sqrt{k!}}(k-1)k,
            \,\,\,\,\,\text{if } \,\,k \ge 2 
	\end{cases}
\tag{S6}
\end{equation}

\!\!\!\!\!\!In addition, the factor $b_k^{(2)}$ is also proportional to the ratio of the normalization factors of the measurement induced $CV$ states of a certain parity. The analytical form of these normalization factors is determined by the peculiarities of the $CV$ states of a certain parity. The normalization factors are polynomials in the derivative of the analytic function $Z(y_1 )=1⁄\sqrt{1-4y_1^2}$ whose even and odd derivatives are determined as $Z^{(2m)}(y_1)=dZ^{2m}⁄dy_1^{2m}$ and $Z^{(2m+1)}(y_1)=dZ^{(2m+1)}⁄dy_1^{(2m+1)}$. 
\par For the measurement induced $CV$ states with superscript (0) we have even ones with  $k=2m$

\begin{equation}\label{eq:S7}
    \ket{\Psi_{2m}^{(0)}(y_1)} =
    \frac{1}{\sqrt{Z^{(2m)}(y_1)}}\sum_{n=0}^{\infty} \frac{y_1^{n}}{\sqrt{(2n)!}}\frac{(2(n+m))!}{(n+m)!}\ket{2n},  
\tag{S7}
\end{equation}

\!\!\!\!\!\!\!and odd states with $k=2m+1$ 

\begin{equation}\label{eq:S8}
    \ket{\Psi_{2m+1}^{(0)}(y_1,B)} =
    \sqrt{\frac{y_1}{{Z^{(2m+1)}}(y_1)}}\sum_{n=0}^{\infty} \frac{y_1^{n}}{\sqrt{(2n+1)!}}\frac{(2(n+m+1))!}{(n+m+1)!}\ket{2n+1}.  
\tag{S8}
\end{equation}

\!\!\!\!\!\!\!For the $CV$ states with superscript (2) we have

\begin{equation}\label{eq:S9}
    \ket{\Psi_{0}^{(2)}(y_1)} =
    \frac{1}{\sqrt{{G_0^{(2)}}(y_1,B)}}\sum_{n=0}^{\infty} \frac{y_1^{n}}{\sqrt{(2n)!}}\frac{(2n)!}{n!}n\ket{2n},  
\tag{S9}
\end{equation}

\!\!\!\!\!\!\!for $k=0$

\begin{equation}\label{eq:S10}
    \ket{\Psi_{1}^{(2)}(y_1,B)} =
    \sqrt{\frac{y_1}{{G_0^{(2)}}(y_1,B)}}\sum_{n=0}^{\infty} \frac{y_1^{n}}{\sqrt{(2n+1)!}}\frac{(2n)!}{n!}(2n+1)(1-Bn)\ket{2n+1}  ,
\tag{S10}
\end{equation}

\!\!\!\!\!\!\!for $k=1$  

\begin{equation}\label{eq:S11}
    \ket{\Psi_{2m}^{(2)}(y_1,B)} =
    \frac{1}{\sqrt{{G_{2m}^{(2)}}(y_1,B)}}\sum_{n=0}^{\infty} \frac{y_1^{n}}{\sqrt{(2n)!}}\frac{(2(n+m-1))!}{(n+m-1)!}f_{2n\,2m}^{(2)}(B)\ket{2n}  ,
\tag{S11}
\end{equation}

\!\!\!\!\!\!\!for $k=2m$ containing the additional inner amplitude with two subscripts $2n$ and $2m$, i.e. 

\begin{equation}\label{eq:S12}
    f_{2n\,2m}^{(2)}(B) = 1 - \frac{2B}{2m-1}2n+\frac{B^2}{(2m-1)2m}(2n-1)2n,
\tag{S12}
\end{equation}

\!\!\!\!\!\!\!which depends on $B$ and

\begin{equation}\label{eq:S13}
    \ket{\Psi_{2m+1}^{(2)}(y_1,B)} =
    \sqrt{\frac{y_1}{{G_{2m+1}^{(2)}}(y_1,B)}}\sum_{n=0}^{\infty} \frac{y_1^{n}}{\sqrt{(2n+1)!}}\frac{(2(n+m))!}{(n+m)!}f_{2n+1\,2m+1}^{(2)}(B)\ket{2n+1},  
\tag{S13}
\end{equation}

\!\!\!\!\!\!\!for $k=2m+1$ with the inner function dependent on $B$

\begin{equation}\label{eq:S14}
    f_{2n+1\,2m+1}^{(2)}(B) = 1 - \frac{2B}{2m}(2n+1)+\frac{B^2}{2m(2m+1)}2n(2n+1).
\tag{S14}
\end{equation}

\!\!\!\!\!\!\!Their normalization factors are the polynomials with derivatives of the function $Z(y_1)$

\begin{equation}\label{eq:S15}
    G_{k}^{(2)}(y_1,B) =
    \begin{cases}
            \frac{1}{4}\Big(y_1\frac{d}{dy_1}\Big(y_1Z^{(1)}(y_1)\Big)\Big), 
            \,\,\,\,\text{if}\,\,\, k = 0 \\[3ex]
			\sum_{l=2}^{4}A_{1l}^{(2)}\Big(y_1\frac{d}{dy_1}\Big)^{l-1}\Big(y_1Z(y_1)\Big) , & 
            \!\!\!\!\!\!\!\!\!\!\!\!\!\!\!\!\!\!\!\!\!\!\!\!\!\!\!\!\!\!\!\!\!\!\!\!\!\!\!\!\!\!\!\!\!\!\!\text{if } k = 1 \\[3ex]
			Z^{(k-2)}(y_1) + \sum_{l=1}^{4}A_{kl}^{(2)}\Big(y_1\frac{d}{dy_1}\Big)^{l-1}\Big(y_1Z^{(k-1)}(y_1)\Big), &\!\!\text{if } k > 0 
	\end{cases}
\tag{S15}
\end{equation}

\!\!\!\!\!\!\!which are formed using the following matrix elements 

\begin{equation}\label{eq:S16}
    A_{kl}^{(2)}(B) =
    \begin{cases}
            A_{12}^{(2)} = (1 + \frac{B}{2})^2,\,\,\,
            A_{13}^{(2)} = -B(1 + \frac{B}{2}),\,\,\,
            A_{14}^{(2)} = \frac{B^2}{4}\,,
            \,\,\,\,\text{if}\,\,\, k = 0 \\[3ex]
			A_{k1}^{(2)} = -\frac{4B}{k-1}(1 + \frac{B}{2k}),\,\,\,
            A_{k2}^{(2)} = \frac{4B^2}{(k-1)^2}(1 + \frac{B^2}{4k^2} + \frac{k-1}{2k} +\frac{B}{k}) \,,\,\text{if} \,\,k = 1 \\[3ex]
			A_{k3}^{(2)} = -\frac{4B^3}{k(k-1)^2}(1 + \frac{B}{2k}),\,\,\,
            A_{k4}^{(2)} = \frac{B^4}{k^2(k-1)^2}\,,\,
            \text{if} \,\,k > 1
	\end{cases}
\tag{S16}
\end{equation}

\par In addition to the given quantities, the measurement induced CV state of a certain parity (\ref{eq:S3}) also contains its own normalization factor $G_k^{(02)}(y_1,B,\varphi)$. Knowledge of the normalization factor $G_k^{(2)}(y_1,B)$ and amplitudes $b_2b_k^{(2)}$ allows us to estimate it as

\begin{equation}\label{eq:S17}
    G_k^{(02)}(y_1,B) = 1 + b_2^2b_k^{(2)2} + \frac{2b_2b_k^{(2)}J_k^{(02)}\cos{\varphi}}{\sqrt{Z^{(k)}(y_1)G_k^{(2)}(y_1,B)}},
\tag{S17}
\end{equation}

where the non-zero cross term between $\ket{\Psi_{k}^{(0)}(y_1,B)}$ and  $\ket{\Psi_{k}^{(2)}(y_1,B)}$ is given by 

\begin{equation}\label{eq:S18}
\begin{split}
J_k^{(02)}(y_1,B) &= \sqrt{Z^{(k)}(y_1)G_{k}^{(2)}(y_1,B)}{\bra{\Psi_{k}^{(0)}(y_1,B)}\ket{\Psi_{k}^{(2)}(y_1,B)}}= \\
& \begin{cases}
            \,\,\,\,\,\,\,\,\,\,\,\,\,\,\,\,\,\,\,\,\,\,\,\,\,\,\,\,\,\,\,\,\,\,\,\,\,\,\,\,\,\,\,\,\,\frac{y_1Z^{(1)}(y_1)}{2}, 
            \,\,\,\,\text{if}\,\,\, k = 0 \\[3ex]
           \,\,\,\,\,\,\,\,\,\,\,\,\,\,\,\,\,\,\,\,\,\,\,\,\,\,\,\,\,\, 2\Big(\sum_{l=2}^{3}A_{1l}^{(02)}\Big(y_1\frac{d}{dy_1}\Big)^{l-1}\Big(y_1Z(y_1)\Big)\Big) , & 
            \,\,\,\,\,\,\,\,\,\,\,\,\,\,\,\!\!\!\!\!\!\!\!\!\!\!\!\!\!\!\!\!\!\!\!\!\!\!\!\!\!\!\!\!\!\!\!\!\!\!\!\!\!\!\!\!\!\!\text{if } k = 1 \\[3ex]
            2\Big(A_{k0}^{(02)}Z^{(k-2)}(y_1) + \sum_{l=1}^{3}A_{kl}^{(02)}\Big(y_1\frac{d}{dy_1}\Big)^{l-1}\Big(y_1Z^{(k-1)}(y_1)\Big)\Big), &\!\!\text{if } k > 1
       \end{cases}
\end{split}
\tag{S18}
\end{equation}

with the coefficients

\begin{equation}\label{eq:S19}
    A_{kl}^{(02)}(B) =
    \begin{cases}
            A_{12}^{(02)} = 1 + \frac{B}{2},\,\,\,
            A_{13}^{(02)} = -\frac{B}{2}\,,
            \,\,\,\,\text{if}\,\,\, k = 0 \\[3ex]
			A_{k0}^{(02)} = k-1,\,\,\,
            A_{k1}^{(02)} = 1 - 2B - \frac{B^2}{k},\,\text{if} \,\,k = 1 \\[3ex]
			A_{k2}^{(02)} = \frac{B^2}{k} + \frac{2B}{k-1} +\frac{B^2}{k(k-1)},\,\,\,
            A_{k3}^{(02)} = \frac{B^2}{k(k-1)}\,,\,
            \text{if} \,\,k > 1
	\end{cases}
\tag{S19}
\end{equation}

\!\!\!\!\!\!Since the output measurement induced $CV$ state of definite parity contains a phase factor $\exp(i\varphi)$ and due to the fact that the $CV$ states of a certain parity  $\ket{\Psi_{k}^{(0)}}$ and  $\ket{\Psi_{k}^{(2)}(y_1)}$ are not orthogonal to each other, then, its normalization term also depends on $\sin \varphi$. 
\par The probability of detecting a particular measurement outcome and, as a consequence, generating the measurement-induced states in equation (\ref{eq:S3}) follows directly from the parameters used

\begin{equation}\label{eq:S20}
    P_k^{(02)}(\varphi) = \frac{c_k^{(0)2}(y_1,B)Z^{(k)}(y_1)}{N_2\cosh s}G_k^{(02)}(y_1,B,\varphi) =\frac{\sqrt{1-4y_1^2(1+B)^2}(y_1B)^kZ^{(k)}(y_1)}{N_2k!} G_k^{(02)}(y_1,B,\varphi).
\tag{S20}
\end{equation}

\!\!\!\!\!\!The distribution is normalized regardless of the values $y_1$, $B$,$\varphi$ and $b_2$, i.e. $\sum_{k=0}^{\infty}P_k^{(02)}(y_1,B,\varphi) = 1$.

\section*{Supplementary note 2: Statistical characteristics CV states (\ref{eq:S3})}

Here we consider analytical expressions for the statistical characteristics inherent in the measurement induced $CV$ states of a certain parity in equation (\ref{eq:S3}). Deriving analytical expressions for the normalization factors allows us to derive an analytical expression for the average number of photons in the state (\ref{eq:S3}) \cite{chap2:2}

\begin{equation}\label{eq:S21}
    \langle n^{(02)}_{k}\rangle = \frac{1}{G_k^{(02)}}\Bigg(
    \frac{\bigl(y_1\frac{d}{dy_1}\bigl)Z^{(k)}}{Z^{(k)}} + 
    b_2^2b_k^{(2)2}\frac{\bigl(y_1\frac{d}{dy_1}\bigl)G^{(2)}_k}{G^{(2)}_k} + 
    2b_2b_k^{(2)}\frac{\bigl(y_1\frac{d}{dy_1}\bigl)J_k^{(02)}}{\sqrt{Z^{(k)}G_k^{(2)}}}\cos \varphi \Bigg),
    \tag{S21}
\end{equation}

\!\!\!\!\!\!where $\langle n \rangle$ indicate averaging over the state (\ref{eq:S3}), the use of which allows us to estimate the rate of its change with change of $\varphi$

\begin{equation}\label{eq:S22}
    \frac{d\langle n^{(02)}_{k}\rangle}{d\varphi} =
    \langle n^{(02)}_{k}\rangle_{\varphi} = 
    \frac{2b_2b_k^{(2)}\sin\varphi}{G_k^{(02)}(\varphi)\sqrt{Z^{(k)}G_k^{(2)}}}\Big(J_k^{(02)}\langle n^{(02)}_{k}\rangle - \Big(y_1\frac{d}{dy_1}\Big)J_k^{(02)}\Big).
    \tag{S22}
\end{equation}

\!\!\!\!\!\!Following the same method one can derive an expression for the noise variance $\triangle n^{(02)2}_k$ = ${\langle n^{(02)2}_{k}\rangle-\langle n^{(02)}_{k}\rangle^2}$, where

\begin{equation}\label{eq:S23}
    \langle n^{(02)2}_{k}\rangle = \frac{1}{G_k^{(02)}(\varphi)}\Bigg(
    \frac{\bigl(y_1\frac{d}{dy_1}\bigl)^2Z^{(k)}}{Z^{(k)}} + 
    b_2^2b_k^{(2)2}\frac{\bigl(y_1\frac{d}{dy_1}\bigl)^2G^{(2)}_k}{G^{(2)}_k} + 
    2b_2b_k^{(2)}\frac{\bigl(y_1\frac{d}{dy_1}\bigl)^2J_k^{(02)}}{\sqrt{Z^{(k)}G_k^{(2)}}}\cos \varphi \Bigg)
    \tag{S23}
\end{equation}

\par An analytical approach through normalization factors $Z^{(k)} $, $G_k^{(2)}$, $J_k^{(02)}$ and the additional multiplier $b_k^{(2)}$ can be used to calculate the quantum Fisher information ($QFI$) $F = 4 \Big ({\bra{\Psi_{\varphi}}\ket{\Psi_{\varphi}}}-\Big\lvert{\bra{\Psi}\ket{\Psi_{\varphi}}}\Big\rvert^2\Big)$. In the terms for the state (\ref{eq:S3}) it becomes

\begin{equation}\label{eq:S24}
    F_{\Psi_k^{(02)}} = 4 X_k^{(02)2}\Big(R_k^{(02)2}\Big(1- \frac{b_2^2b_k^{(2)2}}{G_k^{(02)}}\Big)-2\frac{b_2b_k^{(2)}R_k^{(02)}}{\sqrt{G_k^{(02)}}}\cos \varphi - 1\Big)
    \tag{S24}
\end{equation}

\!\!\!\!\!\!where two new parameters are introduced, obtained from those used previously

\begin{equation}\label{eq:S25}
    R_k^{(02)} = \frac{\sqrt{Z^{(k)}G_{k}^{(2)}G_{k}^{(02)}}}{J_{k}^{(02)}}, 
    \tag{S25}
\end{equation}

\begin{equation}\label{eq:S26}
    X_k^{(02)} = \frac{b_2b_k^{(2)}J_k^{(02)}}{G_k^{(02)}\sqrt{Z^{(k)}G_k^{(2)}}} = \frac{b_2b_k^{(2)}}{R_k^{(02)}\sqrt{G_k^{(02)}}}. 
    \tag{S26}
\end{equation}

\section*{Acknowledgments}
The work of MSP and SAP was supported by the Foundation for the Advancement of Theoretical Physics and Mathematics “BASIS” (Project № 24-1-1-87-1).

\renewcommand{\refname}{References}

\end{document}